 \documentclass[final,5p,times,twocolumn,authoryear]{elsarticle}

\usepackage{amssymb}
\usepackage{lipsum}

\journal{High Energy Astrophysics}
\usepackage{hyperref}
\hypersetup{
    colorlinks=true,
    linkcolor=myblue,
    citecolor=myblue,
    urlcolor=myblue
}
\usepackage{graphicx}
\usepackage{amsmath}
\usepackage[pagewise]{lineno}
\usepackage{newtxtext,newtxmath}
\usepackage{xcolor}
\usepackage{multirow}
\usepackage{mwe}
\usepackage{comment}
\usepackage{adjustbox}
\usepackage{graphicx}
\usepackage{amsmath}
\usepackage{threeparttable}
\usepackage{footnote}
\usepackage{tabularx}
\usepackage{graphicx}
\usepackage{soul}
\usepackage{epstopdf}
\usepackage{footnote}
\usepackage{float}
\usepackage{pdfpages}

\usepackage[utf8]{inputenc} 
\usepackage[T1]{fontenc} 

\begin{document}

\begin{frontmatter}



\title{Study on the magnetic field of the ultraluminous X-ray pulsar RX
J0209.6-7427}


\author[1]{Amar Deo Chandra}
\ead{amar.deo.chandra@gmail.com}
\address[1]{Aryabhatta Research Institute of Observational Sciences, Manora Peak, Nainital, Uttrakhand, 263001, India}

\begin{abstract}
RX J0209.6-7427 is an ultraluminous X-ray pulsar (ULXP) having spin period of about 9.3 s.
To date, no cyclotron resonance scattering features have been detected in this source, which can enable direct measurement of the magnetic field of the pulsar. We estimate the surface magnetic field of the neutron star in this source using different models and find that the inferred magnetic field lies in the range of $2.4-4 \times 10^{13}$G. We study the magnetic field and spin period evolution of the source using existing models and find that the magnetic field will decay to about  $\sim 10^{9}$G assuming steady accretion and the source will become a millisecond pulsar at the end of the accretion phase of the accreting binary. Comparison between the magnetic field and the spin period of other ULXPs with those of magnetars suggests that some ULXPs may have magnetar-like strong dipolar magnetic fields. Studying the magnetic and spin period evolution of ULXPs may
be helpful for understanding magnetar evolution and the millisecond pulsar formation.
\end{abstract}


\end{frontmatter}




\section{Introduction}
\label{introduction}

Ultraluminous X-ray sources (ULXs) were first detected as extragalactic sources having isotropic X-ray luminosity exceeding the Eddington limit
for a stellar-mass ($\sim 10~\rm M_{\odot}$)  black hole \citep{kaaret2017ultraluminous}. However, this paradigm was challenged with the discovery of coherent pulsations from a few ULXs confirming that some ULXs may be powered by accretion onto neutron stars \citep{bachetti2014,furst2016,israel2017a,israel2017b,weng2017,townsend2017,tsygankov2017smc, carpano2018,wilson2018,castillo2019,sathyaprakash2019,chandra2020study}. The underlying mechanism which powers super-Eddington luminosities in these ultraluminous X-ray pulsars (ULXPs) are not well understood. It is surmised that ULXPs may have strong magnetic fields (B$\sim 10^{14} ~G$) which can diminish the electron scattering cross-section and power super-Eddington luminosities \citep{canuto1971thomson,basko1976limiting,paczynski1992gb,dall2015nustar,ekcsi2015ultraluminous,mushtukov2015maximum,tong2015accreting}. The high magnetic field in these objects may have multipolar magnetic field components \citep{israel2017b,brice2021super}. It is also likely that the super-Eddington luminosities in ULXPs may be supported by high accretion rates abetted with collimated tight beaming \citep{king2016ulxs}.

The magnetic field plays a crucial role in the observational features and evolution of ULXs harbouring neutron stars. There are several methods to estimate the dipole magnetic fields of neutron stars. The accretion torque acting on the neutron star can be estimated from its spin evolution, and this torque is dependent on the interaction between the accretion disc and the dipole magnetic field \citep{ghosh1979accretion,kluzniak2007magnetically}. The transition in the pulse profile of an accreting pulsar from a fan beam to a pencil beam or vice-versa is related to the critical luminosity, which is dependent on the magnetic field strength of the pulsar \citep{becker2012spectral,mushtukov2015critical,chandra2023astrosat}. The magnetic field can also be estimated using the torque balance equation and the minimum luminosity on the propeller line \citep{christodoulou2016x}. The dipole magnetic field of the pulsar can be inferred directly from the detection of cyclotron resonance scattering features (CRSFs) in the X-ray spectra of accreting pulsars \citep{trumper1978evidence,wheaton1979absorption,kendziorra1994evidence,mihara1995observational,coburn2002magnetic,makishima2003measuring,staubert2019cyclotron,chandra2023astrosat}.

RX J0209.6-7427 is a transient Be/X-ray binary pulsar first detected in the \textit{ROSAT} all-sky survey during the 1993 outburst \citep{kahabka2005discovery}. The source lay dormant for nearly three decades until 2019, when it underwent a giant outburst exhibiting 9.3 s pulsations \citep{iwakiri2019atel}. The inferred peak X-ray luminosity during the outburst exceeded the Eddington luminosity and reached $1.6 \pm 0.1 \times 10^{39}$\,ergs\,s$^{-1}$ for a distance of 60 kpc for the source \citep{chandra2020study}, which confirmed that RX J0209.6-7427 belonged to the class of transient ULXPs. The magnetic field of the neutron star in this source is unknown as no
CRSFs have been detected in the outburst spectra of this source from \textit{AstroSat} \citep{chandra2020study}, \textit{NuStar} \citep{vasilopoulos2020} and \textit{Insight-HXMT}  \citep{hou2022fan} observations during the 2019 outburst of this source. It was suggested from the non-detection of CRSFs in this source that any CRSFs would lie beyond the 80 keV energy band \citep{chandra2020study,hou2022fan}. However, \citet{chandra2020study} inferred the surface magnetic field of the neutron star to be
$\sim 2.8 \times 10^{13} ~G$ using the torque balance equation and the minimum luminosity on the propeller line \citep{christodoulou2016x,chandra2025long}. The surface magnetic field was estimated to be $\sim 3 \times 10^{12} ~G$ (possibly underestimated by a factor of about 2) from the spin-up of the pulsar during the 2019 outburst \citep{vasilopoulos2020}. Using critical luminosity of about $3 \times 10^{38}$\,ergs\,s$^{-1}$, the surface magnetic field was inferred to be $\sim 2 \times 10^{13} ~G$ \citep{hou2022fan}. A similar surface magnetic field of about $2 \times 10^{13} ~G$ was estimated using the critical luminosity inferred from the transition detected in the pulse profile from a fan beam to a pencil beam \citep{liu2022comparing}. Using the model suggested by \citet{mushtukov2015critical} which accounts for the different polarisation modes of the emitted X-ray photons, the lower limit of the energy of the CRSF was estimated to be about 200 keV, which
suggested that the minimum surface magnetic field of the neutron star should be $\sim 2.2 \times 10^{13} ~G$ \citep{hou2022fan}.

In this paper, we estimate the surface dipole magnetic field of RX J0209.6-7427 using different accretion models. After introduction, in section 2 we describe the various models and infer the surface magnetic fields. In section 3, we simulate the decay of the dipole magnetic field on long timescales. We study the evolution of the magnetic field and spin period of the pulsar and compare its evolutionary track with known binary pulsars, magnetars and other confirmed ULXPs. We discuss and summarise
our findings in sections 4 and 5 respectively.

\section{Magnetic field calculation}
The neutron star in RX J0209.6-7427 accretes matter from the companion star via an accretion disc during outbursts  as the pulsar exhibited a rapid spin-up during the 2019 outburst having a spin-up rate of $\sim 1.75 \times 10^{-8}$ \,s \,s$^{-1}$ \citep{chandra2020study}. The accretion process is strongly dependent on the magnetospheric
(Alfv\'{e}n) radius ($r_{\rm m}$) which is given by \citep{davidson1973neutron},

\begin{equation}
r_{\rm m}=\xi\left(\frac{\mu^{4}}{2GM\dot{M}^{2}}\right)^{1/7}, \label{eq1}
\end{equation}

where $\xi$ is a dimensionless parameter ($\xi \sim 1$), $\mu=B_{\rm s}R^{3}$ where $B_{\rm s}$ and R are the surface dipole magnetic field and radius of the neutron star respectively, $G$ is the gravitational constant, $M$ is the mass of the neutron star and $\dot{M}$ is the accretion rate at the Alfv\'{e}n radius. Using $R=10^6$ cm and $G=6.67\times 10^{-8}$~cm$^3$~g$^{-1}$~s$^{-2}$ we obtain,

\begin{equation}
r_{\rm m}=1.6\times 10^{8}\xi \dot{M}^{-2/7}_{18}M^{-1/7}_{1.4}\mu^{4/7}_{30}~\rm cm,
\end{equation}
where $\dot{M}_{18}$, $M_{1.4}$ and $\mu_{30}$ are in units of $10^{18}~\rm g\,s^{-1}$, $1.4~\rm M_{\odot}$, and $10^{30}~\rm G\,cm^{3}$ respectively.

The neutron star can be spun up during outbursts by angular momentum imparted by the accreted material from the donor star. In addition, another spin-up torque can be attributed to the magnetic coupling between the neutron star and the inner portion of the accretion disc \citep{ghosh1979accretion}. The spin change in the neutron star due to magnetic torques is dependent on the fastness parameter
$\omega_{\rm s}=\Omega_{NS}/\Omega(r_{\rm m})$ where $\Omega_{NS}$, and $\Omega(r_{\rm m})$ are the angular velocity of the neutron star, and the Keplerian angular velocity at the magnetospheric radius respectively. For $\omega_{\rm s} < 1$, the neutron star is spun-up by the accreted matter, while for $\omega_{\rm s} > 1$  the neutron star is spun-down due to the magnetic interaction between the disc and the neutron star. Ignoring the contribution of the magnetic torque we obtain \citep{chen2017constraining},

\begin{equation}
-\frac{2\pi I\dot{P}}{P^{2}}\leq \dot{M}\sqrt{GMr_{\rm m}}, \label{eq2}
\end{equation}
where P, $\dot{P}$ and $I=\frac{2MR^2}{5}$ are the spin period, spin-up rate and the moment of inertia of the neutron star respectively. Using $r_{\rm m}$ from equation \ref{eq1} in equation \ref{eq2} and $I=1.92 \times 10^{45}\rm \,g\,cm^3$, the lower limit on the dipole magnetic moment ($\mu_{\rm min,30}$) is obtained as,

\begin{equation}
\mu_{\rm min,30}\ge 8.96 \times 10^{37} \xi^{-7/4}\dot{M}^{-3}_{18}M^{-3/2}_{1.4} P^{-7} \dot{P}^{7/2}~\rm G\,cm^{3}.
\end{equation}

Using P=9.3 s and $\dot{P} \sim 1.75 \times 10^{-8}$ \,s \,s$^{-1}$ \citep{chandra2020study} for RX J0209.6-7427, the minimum dipole moment is given by,

\begin{equation}
\mu_{\rm min,30}=1.06 \times 10^{4} \xi^{-7/4}\dot{M}^{-3}_{18}M^{-3/2}_{1.4}~\rm G\,cm^{3}. \label{eq5}
\end{equation}

The maximum limit on the magnetic dipole moment can be obtained by equating the magnetospheric radius and the co-rotation radius of the neutron star ($r_{\rm co}$), which is given by,

\begin{equation}
r_{\rm co} =\biggl( {\frac{GMP^{2}}{4\pi^{2}}} \biggr)^{1/3}~\rm cm. \label{eq6}
\end{equation}

From equations \ref{eq1} and \ref{eq6} and assuming $r_{\rm m}=r_{\rm co}$, the maximum dipole magnetic moment ($\mu_{\rm max,30}$) is obtained as,

\begin{equation}
\mu_{\rm max,30}\le 1.09  \xi^{-7/4}\dot{M}^{1/2}_{18}M^{5/6}_{1.4} P^{7/6} ~\rm G\,cm^{3}. \label{eq7}
\end{equation}

Equating $\mu_{\rm max,30} \simeq \mu_{\rm min,30}$ and using M=$1.4~\rm M_{\odot}$, the minimum accretion rate at the magnetospheric radius is estimated as

\begin{equation}
\dot{M}_{\rm min}= 6.6\times 10^{18}~{\rm g\,s^{-1}}.
\end{equation}

The estimated minimum accretion rate $\dot{M}_{\rm min}$ is about 6.6 times the Eddington accretion rate $\dot{M}_{\rm Edd}$ ($\sim 1.0\times 10^{18}~\rm g\,s^{-1}$) for spherical accretion onto a neutron star. This implies that the source is likely to undergo super-Eddington accretion during outbursts. In fact, the inferred luminosity in the soft X-ray band (0.1-2.4 keV) during the 1993 outburst was $\sim 1.0\times 10^{38}~{\rm erg~s}^{-1}$ \citep{kahabka2005discovery}, suggesting broadband super-Eddington accretion. Similarly, the pulsar underwent super-Eddington accretion during the 2019 outburst \citep{chandra2020study,vasilopoulos2020}.

The geometry of the emission region depends on the mass accretion rate onto the neutron star. At low mass accretion rates known as the sub-critical regime, the accretion rate is $\lesssim 10^{17}~\rm g\,s^{-1}$ and in this case, the accreted matter reaches the surface of the neutron star and forms hot spots on the surface of the neutron star \citep{zel1969x}. At higher accretion rates known as the super-critical regime, the accretion rate is $\gtrsim 10^{17}~\rm g\,s^{-1}$ and the accretion dynamics is impacted by radiation pressure and the accreted matter is decelerated in the radiation-dominated shock above the surface of the neutron star \citep{basko1976limiting,wang1981plasma,mushtukov2015maximum,abolmasov2023simulating}. In the super-critical regime, an accretion column is formed above the surface of the neutron star, which is supported by the radiation pressure and confined by the strong magnetic field of the neutron star \citep{shapiro1983black}. The maximum accretion rate onto the neutron star in the case of formation of an accretion column is given by \citep{basko1976limiting,chen2017constraining},

\begin{equation}
\dot{M}_{\rm max}= \frac{l_{0}}{2\pi d_{0}}\dot{M}_{\rm Edd}\simeq 6.4\times 10^{18}\Bigg(\frac{l_{0}/d_{0}}{40}\Bigg)M_{1.4}~\rm g\,s^{-1} , \label{eq9}
\end{equation}

where $l_{0}$, and $d_{0}$ are the length and the thickness of the accretion column, respectively. The cross-section of the accretion column is akin to a narrow circular ring and the radius of this ring on the neutron star's surface ($r_{\rm column}$) is given by \citep{chen2017constraining},

\begin{equation}
r_{\rm column}= R\sqrt{\frac{R}{r_{\rm m}}}.
\end{equation}

The cross-section radius of the accretion column is weakly dependent on the magnetic field ($\propto B^{-2/7}$) and the accretion rate ($\propto \dot{M}^{1/7}$). Using $\xi \sim 1$, $\dot{M}_{18}=6.6$, $M_{1.4}=1$, and $\mu_{30}\sim 3$ \citep{vasilopoulos2020}, the magnetospheric radius is $\sim 1.75\times 10^{8}~\rm cm$. Using $R=10^6$ cm and $r_{\rm m} \sim 1.75\times 10^{8}~\rm cm$, the radius of the accretion column ($r_{\rm column}$) is $\sim 7.56\times 10^{4}~\rm cm$. The estimated length of the accretion column ($l_{0}=2\pi r_{\rm column}$) is $\sim 4.75\times 10^{5}~\rm cm$. Assuming a thin accreting wall in the accretion column i.e. $d_{0}\ll r_{\rm column}$, $d_{0}=10^{4}~\rm cm$ \citep{mushtukov2015maximum,chen2017constraining}. Using equation \ref{eq9}, the maximum accretion rate ($\dot{M}_{\rm max}$) is estimated to be $\sim 7.6\times 10^{18}~{\rm g\,s^{-1}}$.

\begin{figure}
\centering
  \includegraphics[width=\columnwidth]{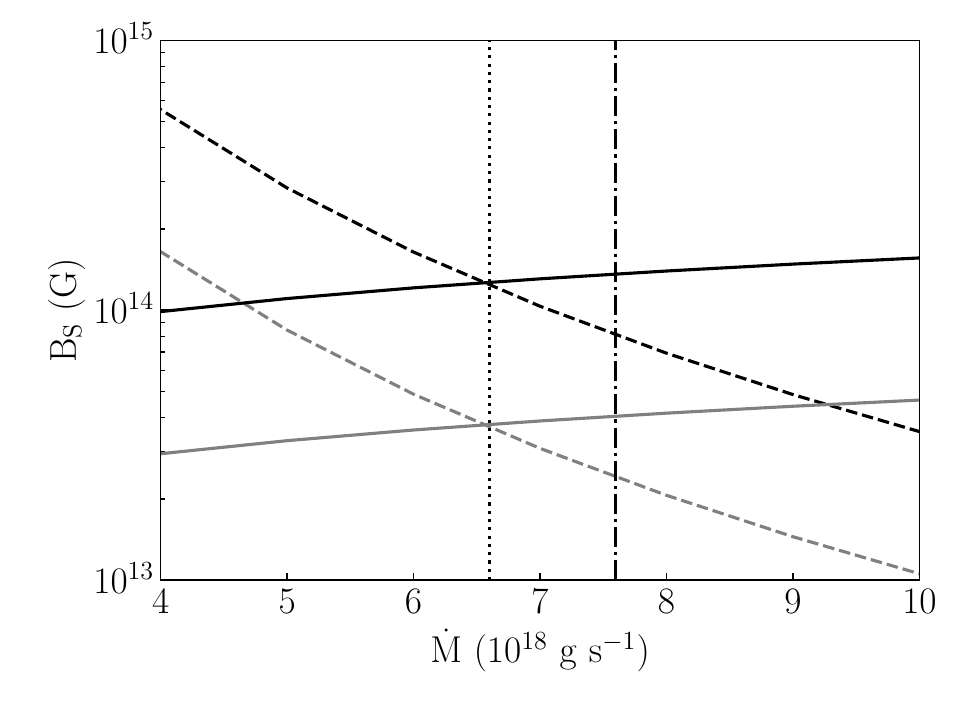}
  \caption{Variation of the surface dipole magnetic field of the neutron
star ($B_S$) and the accretion rate at the magnetospheric radius ($\dot{M}$). The maximum and the minimum surface dipole magnetic fields are marked by solid and dashed curves (with the gray and black curves corresponding to
$\xi$=1.0 and 0.5, respectively), respectively. The vertical dash-dotted and dotted lines denotes
the upper and lower limits of the accretion rate respectively.}
 \label{f1}
\end{figure}

Using equation \ref{eq7}, $\mu=B_{\rm s}R^{3}$, $\xi \sim 1$, $\dot{M}_{18}=7.6$ and P=9.3 s the maximum surface dipole magnetic field of the neutron star ($B_{s,max}$) is estimated to be about $4 \times 10^{13}\,G$. The minimum surface dipole magnetic field of the neutron star ($B_{s,min}$) is estimated to be about $2.4 \times 10^{13}\,G$ using $\xi \sim 1$, and $\dot{M}_{18}=7.6$ in equation \ref{eq5}. The estimated lower and upper limits on the surface magnetic field of the neutron star (Fig. \ref{f1}) is less than the quantum critical limit of the magnetic field $B_Q \equiv m_e^2c^3/(e\hbar)=4.41 \times 10^{13}$~G.

The magnetic field can also be estimated independently by using the torque balance equation \citep{christodoulou2016x},

\begin{equation}
B = \left(2\pi^2 \xi^7\right)^{-1/4}
\sqrt{\frac{G M I}{R^6} |\dot{P_S}|} \, ,
\end{equation}

where $\xi$ is a dimensionless parameter, which is the ratio of the inner edge of the accretion disc and the magnetospheric radius \citep{ghosh1979accretion,wang1996location,christodoulou2016x} and $\dot{P_S}$ is the rate of spin change of the pulsar during outbursts. Using $\xi$=1, $M=1.4 ~M_{\odot}$, R=10 km, $\dot{P_S} \sim 1.75 \times 10^{-8}$ \,s \,s$^{-1}$ \citep{chandra2020study} and $G=6.67\times 10^{-8}$~cm$^3$~g$^{-1}$~s$^{-2}$, the estimated magnetic field of the neutron star is $\sim 2.87 \times 10^{13} ~G$, which lies within the range of surface magnetic fields ($2.4 \times 10^{13}\,G  \leqslant B_s \leqslant 4 \times 10^{13}\,G$) estimated earlier.
The magnetic field can also be estimated independently from the minimum luminosity on the propeller line \citep{christodoulou2016x},

\begin{equation}
B = 8.0\times 10^{11} \sqrt{\frac{L_{X,38}}{\eta}}\left(\frac{P_S}{1~{\rm s}}\right)^{7/6} ~~{\rm G}\, ,
\label{pline4}
\end{equation}

where $L_{X,38} = L_X/1.0\times 10^{38}~{\rm erg~s}^{-1}$. Using $P_S = 9.3$~s, $L_{X,38} \sim 1.8$, and $\eta = 1/4$ \citep{christodoulou2016x}, the estimated magnetic field is $\sim 2.8 \times 10^{13} ~G$, which again lies within the range of magnetic fields estimated earlier using equations \ref{eq5} and \ref{eq7}.

The magnetic field of the neutron star can also be estimated independently using the models given by \cite{ghosh1979accretion} and \cite{kluzniak2007magnetically}, which are applicable to disc-fed systems. The Ghosh and Lamb model \citep{ghosh1979accretion} is applicable to systems irrespective of whether they have achieved spin equilibrium and the model predicts \citep{klus2014spin,chandra2025long}

\begin{equation}
  -\dot{P}=5.0\times10^{-5} \mu_{30}^{2/7} n(\omega_{s}) R_{6}^{6/7}\left(\frac{M}{M_\odot}\right)^{-3/7}I_{45}^{-1} (PL_{37}^{3/7})^2,
  \label{eq13}
\end{equation}

where $\dot{P}$ is the long-term spin derivative of the neutron star (in s yr$^{-1}$) and $n(\omega_{s})$ is the dimensionless accretion torque and
depends on the fastness parameter $\omega_{s}$ \citep{ghosh1979accretion,klus2014spin}. Using $\dot{P}\sim 0.45$ s yr$^{-1}$ \citep{hou2022fan}, $n(\omega_{s})\sim$1, $R=10^6$ cm, $M=1.4 ~M_{\odot}$, $I_{45}$=1.92\,g\,cm$^3$, P=9.3 s and $L \sim 1.6\times 10^{39}$\,erg\,s$^{-1}$ \citep{chandra2020study}, the magnetic field is estimated to be $\sim 9.14 \times 10^{13} ~G$, which is higher by a factor of about 2 than the quantum critical limit of the magnetic field. This estimated magnetic field is also more than that estimated upper limit of the magnetic field using equation \ref{eq7} by a factor of about 2. \citet{meng2022research} estimate magnetic field of $\sim 2 \times 10^{13} ~G$ using the Ghosh and Lamb model and using $\omega_s=0.11$. This estimated magnetic field is also less than the estimated lower limit of the magnetic field using equation \ref{eq5} by a factor of about 1.2.  Using the Kluzniak and Rappaport model \citep{kluzniak2007magnetically}

\begin{equation}
-\dot{P}=8.2\times10^{-5}\mu_{30}^{2/7} g(\omega_{s}) R_{6}^{6/7}\left(\frac{M}{M_\odot}\right)^{-3/7}I_{45}^{-1} (PL_{37}^{3/7})^2,
\end{equation}

where $g(\omega_{s})$ depends on the fastness parameter $\omega_{s}$ and is nearly equal to unity. Using $\dot{P}\sim 0.45$ s yr$^{-1}$ \citep{hou2022fan}, $g(\omega_{s})\sim$1, $R=10^6$ cm, $M=1.4 ~M_{\odot}$, $I_{45}$=1.92\,g\,cm$^3$, P=9.3 s and $L \sim 1.6\times 10^{39}$\,erg\,s$^{-1}$ \citep{chandra2020study}, the magnetic field estimated using this model is $\sim  1.61\times 10^{13} ~G$, which is smaller than that estimated using the Ghosh and Lamb model by a factor of about 6. This estimated magnetic field is also less than the estimated lower limit of the magnetic field using equation \ref{eq5} by a factor of about 1.5.

\section{Magnetic field evolution}
The evolution of the magnetic field of a neutron star during the accretion phase in a neutron star binary system has been explored in several studies \citep{bisnovatyi1974pulsars,taam1986magnetic,shibazaki1989does,geppert1994accretion,van1995magnetic,melatos2000evolution,zhang2006bottom,heuvel2009formation,ho2011evolution,igoshev2021evolution}. We use the model of accretion-induced magnetic field decay of a neutron star proposed by \citet{zhang2006bottom} (hereafter referred to as the ZK model) to study the magnetic field evolution of RX J0209.6-7427. We assume steady accretion with a defined accretion time and rate to study the magnetic field evolution of this source. The ZK model has been used to study the magnetic evolution of NuSTAR J095551+6940.8, NGC 300 ULX1 and NGC 7793 P13 \citep{pan2016magnetic,pan2022study,meng2022research}. In this model, the assumption is that the accreted matter is channelled by the strong magnetic field to the polar cap region, where the magnetic field lines in the polar region are  pushed away towards the equatorial region. In this process, the polar magnetic field gets diminished and the compressed stronger field lines at the equator region mostly get squeezed into the inner crust of the neutron star as a result of which the polar magnetic field decays to the bottom field. The analytic solution of the evolved magnetic field in the ZK model is given by \citep{zhang2006bottom},

\begin{equation}
\label{eq15}
B=\frac{B_{\rm f}}{\{1-[C/{\rm exp}(y)-1]^2\}^{7/4}},
\end{equation}
where $B_{\rm f}$ is the bottom magnetic field ($B_{\rm f}\simeq 1.32\times10^8(\dot M/\dot M_{\rm Edd})^{1/2}m^{1/4}R_6^{-5/4}\phi^{-7/4}\,\rm G$, ), $m=1.4~\rm M_{\odot}$  is the mass of the neutron star, R=10 km is the typical neutron star radius, and $\phi$ is the ratio of the magnetospheric radius to the Alfv$\acute{\rm e}$n radius which is assumed to be 0.5.
$C=1+(1-X_0^{2})^{1/2}\sim 2$ where $X_0^{2}=(B_{\rm f}/B_0)^{4/7}$, $B_0$ is the initial magnetic field of the neutron star at the beginning of evolution. The parameter $y=2\xi\Delta M/7M_{\rm cr}$ is the ratio of the accreted mass ($\Delta M$) to the crust mass ($M_{\rm cr}\sim0.2\,M_{\odot}$) of a neutron star. The mass accreted by a neutron star is given by $\Delta M=\dot M T_{\rm ac}$, where $T_{\rm ac}$ is the accretion time. The efﬁciency factor $\xi$ expresses the frozen ﬂow of the magnetic line due to the plasma instability and is taken as unity for completely frozen field lines.

\begin{figure}
\centering
  \includegraphics[width=\columnwidth]{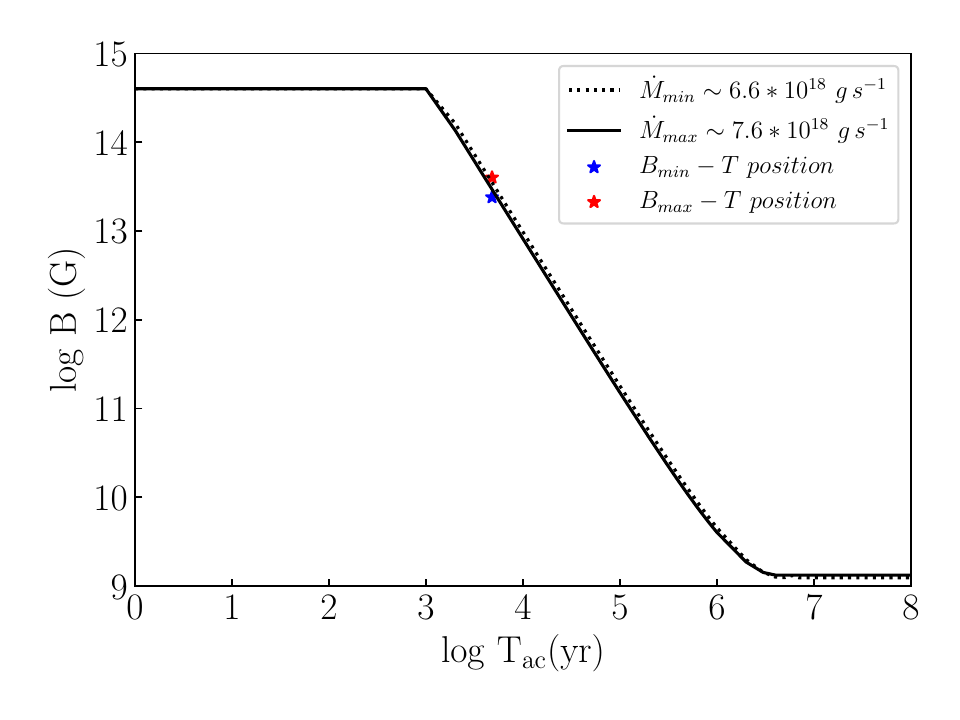}
  \caption{The decay of the magnetic field with the accretion time of RX J0209.6-7427. The evolution is assumed to begin with $T_{ac}=0$ and $B_{0}=4 \times 10^{14}\,G$. The curves plotted with dotted and solid lines show the magnetic field evolution for $\dot{M_{min}}\sim 6.6\times 10^{18}~{\rm g\,s^{-1}}$ and $\dot{M_{max}}\sim 7.6\times 10^{18}~{\rm g\,s^{-1}}$ respectively. The magnetic field decays to the bottom value of $\sim 1.3 \times 10^{9}\,G$ in about $10^{6}$ years. The $B-T_{ac}$ positions are plotted for the minimum and the maximum estimated values of the magnetic field.}
 \label{f2}
\end{figure}

The evolution of the magnetic field is simulated for $\dot{M}_{min}\sim 6.6\times 10^{18}~{\rm g\,s^{-1}}$ and $\dot{M}_{max}\sim 7.6\times 10^{18}~{\rm g\,s^{-1}}$ as shown in Fig. \ref{f2}. We assume that the magnetic evolution begins at $T_{ac}=0$ and $B_{0}=4 \times 10^{14}\,G$. It is observed that the initial magnetic field decays to the present estimated magnetic field of about $2.4-4 \times 10^{13}\,G$ in about $5 \times 10^3$ years. The magnetic field will decay to the bottom magnetic ﬁeld of $\sim 1.3 \times 10^{9}\,G$ in about $6 \times 10^6$ years after accreting mass of about $0.6~\rm M_{\odot}$. The accretion time ($T_{ac}$) of a neutron star in a binary system is related to the lifetime of the companion star in the main sequence, given by \citep{shapiro1983black},

\begin{equation}
T_{\rm ac}=1.3\times10^{10}\, f \,m_{\rm c}^{-2.5}\,\rm yr,
\label{eq16}
\end{equation}
where $m_{\rm c}$ is the mass of the companion star (in units of solar mass) and $f$ is the accretion efficiency factor, which is about 0.1 \citep{shapiro1983black}.
The estimated mass of the companion star in RX J0209.6-7427 is about 11-23 $\rm M_{\odot}$ \citep{coe2020major}. The corresponding accretion time is estimated to be about $0.5-3\times 10^{6}$ years using equation \ref{eq16}. The magnetic field of RX J0209.6-7427 will decay to $\sim 1.4-10 \times 10^{9}\,G$ within this accretion time limited by the mass of the companion star. The accreted mass from the companion star during this duration will be about 0.05-0.3 $\rm M_{\odot}$.

\begin{figure}
\centering
  \includegraphics[width=\columnwidth]{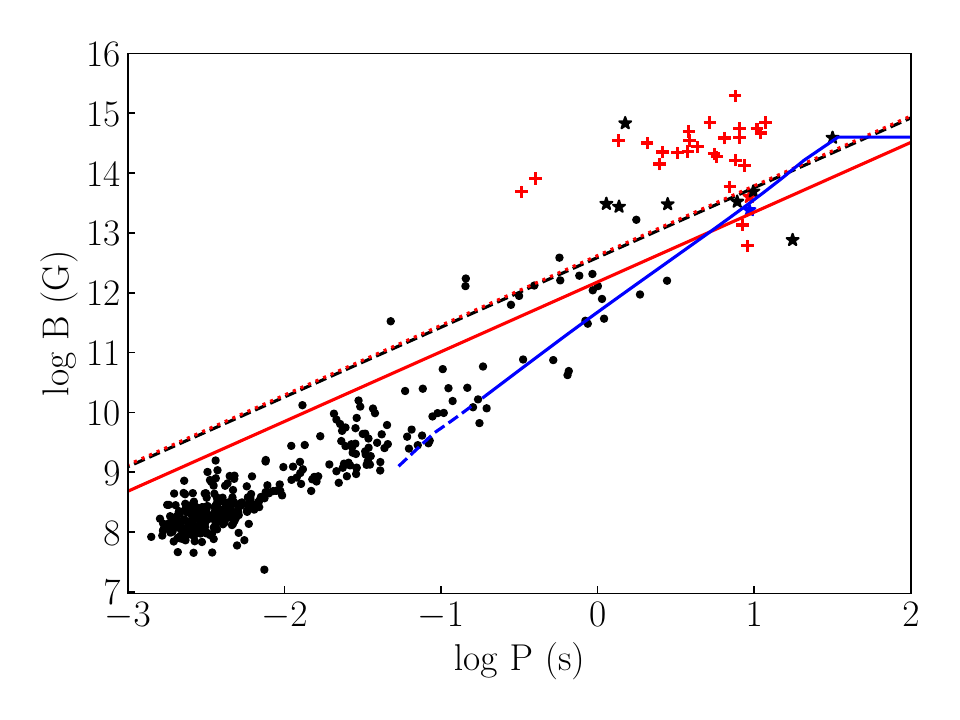}
  \caption{Magnetic field vs spin period evolution of RX J0209.6-7427 is shown by blue curve. The blue and red stars in the figure show the B-P position of RX J0209.6-7427 for $B_{min}=2.4 \times 10^{13}\,G$ and $B_{max}=4 \times 10^{13}\,G$ respectively, having spin period of about 9.3 s. The dotted, dashed and solid lines exhibit the spin-up line having accretion rate $\dot M_{max}\sim 7.6\times 10^{18}~{\rm g\,s^{-1}}$, $\dot{M_{min}}\sim 6.6\times 10^{18}~{\rm g\,s^{-1}}$ and the Eddington accretion rate, respectively. The B-P positions of binary pulsars obtained from the ATNF pulsar catalogue (\url{https://www.atnf.csiro.au/research/pulsar/psrcat/}, \citet{manchester2005australia}) are marked by solid dots while those of magnetars are marked by plus signs. The B-P positions of other confirmed ultraluminous X-ray pulsars M82 X-2, NGC 5907 ULX-1, NGC 300 ULX-1, NGC 1313 X-2, Swift J0243.6+6124, SMC X-3, NGC 2403 ULX and M51 ULX-7 obtained from \citet{meng2022research} are shown by black stars.}
 \label{f3}
\end{figure}

The evolution of the magnetic ﬁeld and the spin period of RX J0209.6-7427 is simulated using the following equation \citep{ye2020exploring}

\begin{equation}
\frac{2\pi}{P}=\frac{2\pi}{P_0}+ \xi^{'} {B^{2/7}_{012}} \Bigg[\Bigg(1+\frac{\dot{M} t}{m_B}\Bigg)^{5/7}-1\Bigg],
\end{equation}

where $P_0$ is the initial spin period of the neutron star, $\xi^{'}=\frac{7}{5}k \Big(\frac{m_B}{\dot{M}}\Big)$ where k is given by $k=1.87\times10^{-10}\dot M_{18}^{6/7}m^{-4/7}R_6^{-8/7}\phi^{1/2}$, $B_{012}$ is the initial dipole magnetic field of the neutron star and $m_B=0.5(B_{\rm f}/B_0)^{4/7}M_{cr}$ \citep{zhang2006bottom}. We assume steady accretion with the defined accretion time and rate. The accretion time is assumed to be the Hubble time $1.3 \times 10^{10}$ yr \citep{ade2016planck}. The accretion rate used is $\dot{M}=\dot{M}_{min}\sim 6.6\times 10^{18}~{\rm g\,s^{-1}}$. We use $P_0=100$ s, $B_{012}$=400 G, m=1.4 $\rm M_{\odot}$, $R=10^6$ cm, $\phi$=0.5, $M_{\rm cr}=0.2\,M_{\odot}$ and $B_f=1.3 \times 10^{9}\,G$.  The evolved B-P path is shown in Fig \ref{f3}, where the dashed part (at the end of the track) is
limited by the accretion time $0.5-3\times 10^{6}$ yr calculated earlier. It is observed that RX J0209.6-7427 will evolve to become a millisecond pulsar having spin period of about hundred milliseconds at the end of the accretion phase.

\section{Discussions}
The magnetic field of other confirmed ultraluminous X-ray pulsars M82 X-2, NGC 5907 ULX-1, NGC 300 ULX-1, NGC 1313 X-2, Swift J0243.6+6124, SMC X-3, NGC 2403 ULX, ULX NGC 7793 P13 and M51 ULX-7 have been estimated in several studies using different models \citep{erkut2020magnetic,chen2021studying,gao2021magnetic,meng2022research}. The estimated magnetic fields lie in the range of $\sim 10^{11-13}$G, $\sim 10^{11-15}$G, $\sim 10^{10-14}$G and  $\sim 10^{13-14}$G in the studies by \citet{erkut2020magnetic}, \citet{chen2021studying}, \citet{gao2021magnetic} and \citet{meng2022research} respectively. This suggests that
ultraluminous X-ray pulsars are likely to have strong magnetic fields. The detection of cyclotron resonance scattering features during outbursts in accreting pulsars can help to reliably measure the magnetic field near the surface of a neutron star. A CRSF around 146 keV has been detected only in one of the ULXPs J0243.6+6124 till date, which yields surface magnetic field of $\sim 1.6 \times 10^{13}$G \citep{kong2022insight}. In fact, the CRSF detected around 146 keV is the highest energy CRSF detected in any accreting pulsar \citep{kong2022insight}. Thus, more detections of CRSFs in ultraluminous X-ray pulsars are required to directly measure the magnetic field of the neutron star in these sources and compare them with those estimated using different models.

There are twelve binary pulsars whose B-P positions
lie above the Eddington spin-up line (Fig \ref{f3}). These pulsars may likely have experienced super-Eddington accretion during their evolution. On comparing the B-P positions of other confirmed ultraluminous X-ray pulsars \citep{meng2022research} with RX J0209.6-7427 (Fig. \ref{f3}), it seems that most of the ULXPs are accreting millisecond pulsars likely having strong magnetic fields ($B\gtrsim 1 \times 10^{13}\,G$). The B-P evolution of RX J0209.6-7427 crosses near the zone of
magnetars and binary pulsars above the Eddington spin-up line and culminates in around the positions of recycled binary pulsars having spin period in the range of milliseconds. Studies of evolution of ULXPs may be helpful to understand magnetar evolution, evolution of accreting millisecond pulsars having strong magnetic fields, and formation of binary pulsars as well as millisecond pulsars both above and below the Eddington spin-up line.

\section{Conclusions}
We estimate the magnetic field strength of RX J0209.6-7427 using different models and find that the estimated field lies in the range of $2.4-4 \times 10^{13}$G. We simulate the magnetic field decay of the neutron star in the binary assuming steady accretion with a defined accretion rate and find that the field will decay to the bottom magnetic field of about $10^{9}$G at the end of the accretion phase. We also study the magnetic field and spin period evolution of the binary confined by the accretion rate and the accretion time, which suggests that the pulsar can evolve to a recycled millisecond pulsar having spin period and magnetic field of about hundred milliseconds and $\sim 10^{9}$G respectively.

\section*{Acknowledgements}
We thank the referee for their constructive suggestions that helped to improve the manuscript. This research has made use of NASA's Astrophysics Data System. ADC acknowledges support from ARIES through post-doctoral fellowship.

\bibliographystyle{elsarticle-harv}
\bibliography{main}






\end{document}